\documentclass[a4paper, 10 pt, conference]{ieeeconf}

\usepackage[utf8]{inputenc}
\usepackage[ruled,vlined]{algorithm2e}
\usepackage{url}
\usepackage[normalem]{ulem}
\usepackage{marginnote}
\usepackage{graphicx}
\usepackage[style=ieee]{biblatex}

\IEEEoverridecommandlockouts                              %
\newcommand{\argmin}{\mathop{\mathrm{argmin}}}

\title{\LARGE \bf Map Matching Algorithm for Large-scale Datasets}

\author{David Fiedler$^{1}$, Michal {\v C}{\' a}p$^{1}$, Jan Nykl$^{2}$, Pavol {\v Z}ileck{\' y}$^{2}$, and Martin Schaefer$^{1}$%
\thanks{$^{1}$Faculty of Electrical Engineering, Artificial Intelligence Center,
        Czech Technical University, Prague, Czech Republic
       {\tt\small david.fiedler@agents.fel.cvut.cz}}%
\thanks{$^{2}$Umotional,
        Prague, Czech Republic
      {\tt\small jan.nykl@umotional.com}}%
}

\begin{document}

\maketitle

\begin{abstract}
GPS receivers embedded in cell phones and connected vehicles generate series of location measurements that can be used for various analytical purposes. A common preprocessing step of this data is the so-called map matching.
The goal of map matching is to infer the trajectory that the device followed in a road network from a potentially sparse series of noisy location measurements.
Although accurate and robust map matching algorithms based on probabilistic models exist, they are computationally heavy and thus impractical for processing of large datasets.
In this paper, we present a scalable map matching algorithm based on Dijkstra's shortest path method, that is both accurate and applicable to large datasets.
Our experiments on a publicly-available dataset showed that the proposed method achieves accuracy that is comparable to that of the existing map matching methods using only a fraction of computational resources.
In result, our algorithm can be used to efficiently process large datasets of noisy and potentially sparse location data that would be unexploitable using existing techniques due to their high computational requirements.

\end{abstract}

\section{Introduction}

With the current spread of GPS receivers, embedded into almost all cell phones and connected cars, a huge amount of GPS data from vehicular traffic is produced. 
The data can be used to analyze various properties of road network like traffic density, travel time or speed, 
or to uncover traffic-related behavioral patterns. 
For all these applications, however, we need to firstly perform the so-called map matching, a process of mapping the GPS records to the road network to obtain the real path driven by the vehicle. 

The map matching problems can be divided into two categories: online map matching and offline map matching. 
The online map matching is a process of determining vehicle path in the road network while the vehicle is driving, the offline map matching computes the path after all GPS measurements have been recorded. 
In this paper, we study the offline map matching problem -- we will use the term ``map matching" to refer to the offline map matching problem in the remainder of this paper.

\begin{figure}
\centering{}
\includegraphics[width=1\columnwidth]{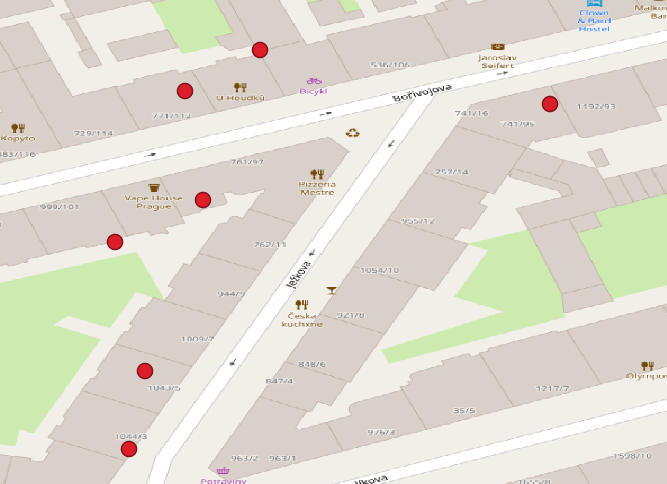}
\caption{\label{fig:bad_records}The measurement noise of a GPS record from Prague}
\end{figure}

\begin{figure}
\centering{}\includegraphics[width=1\columnwidth]{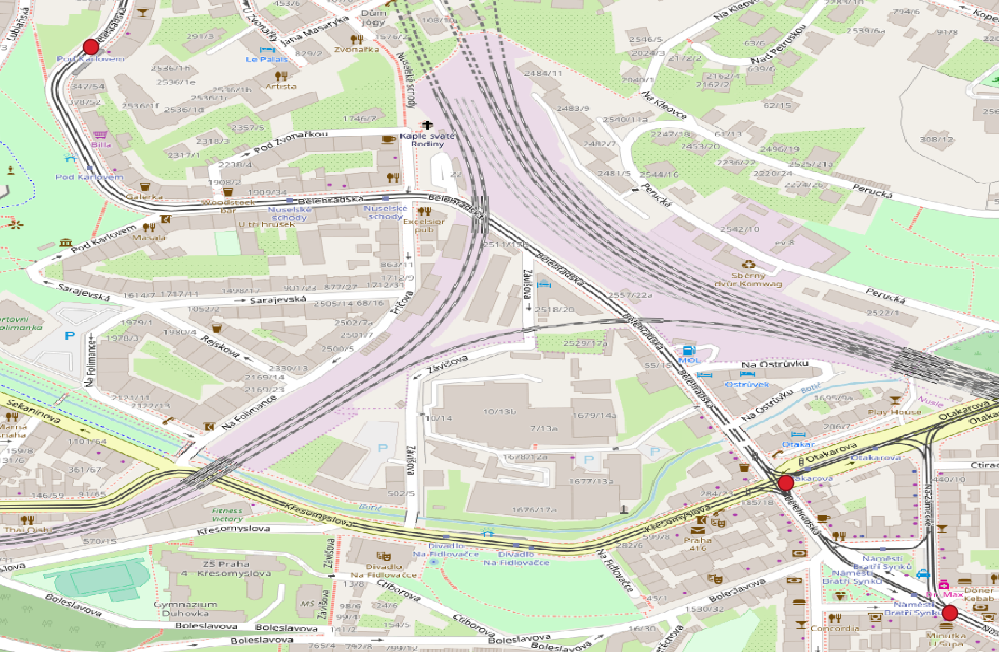}
\caption{\label{fig:bad_records-sparse}The Sparseness of a GPS record from Prague}
\end{figure}

When the location data are sparse and affected by significant measurement errors, the map matching problem becomes challenging because there are typically multiple routes in the road network that could be considered as a match for the measurements.
In an urban environment, in particular, the GPS signal is often affected by multi-path propagation which results in high measurement error. 
See Figure~\ref{fig:bad_records} for an illustration of a typical measurement error in an urban environment.
Furthermore, in order to keep energy consumption low, many devices produce data at low frequency. See Figure~\ref{fig:bad_records-sparse} for an example of a sequence of location measurements produced by a taxi-sharing application running on the driver's mobile phone.

To enable processing of low-frequency high-noise GPS data, a number of map matching algorithms have been proposed.
A typical map matching algorithm is based on the following scheme: firstly, a set of candidate projections (road segments) on the road network are computed for each GPS record, then, the most probable path is chosen so that for each record, there is one candidate projection in the final path.
We refer these algorithms as \emph{global} because they select the path that is globally optimal with respect to some model describing the map matching problem (e.g., the minimal average distance between a measurement and the corresponding candidate edge).
Global map matching algorithms manage to match even very sparse GPS records with high measurements errors.
These algorithms are, however, computationally heavy and thus they are impractical for processing of large datasets, possibly containing millions of GPS traces.
Other methods attempt to optimize the path locally, matching the GPS measurements one by one, we refer to these approaches as \emph{local} map matching algorithms. 
These methods, however, achieve unsatisfactory accuracy, especially on low-frequency high-noise datasets. 
Moreover, the performance of these algorithms is highly sensitive to the choice of parameters of the algorithm, and consequently, these parameters have to be manually tuned for each dataset.

In this paper, we present a global map matching algorithm that achieves accuracy comparable to the state of the art methods using considerably less computation time. 
In result, this method is suitable for processing of large location datasets.
Our solution is based on the forward search in a graph that represents the underlying road network. 
The algorithm performs the graph search in the same manner as Dijkstra's shortest path algorithm.
In contrast to Dijkstra's algorithm, where the edge costs are static and often represent distance or travel time, in the proposed method, the edge costs encode the likelihood of the edge being part of the true route given the observed location measurements and this cost is computed dynamically during the search.

Our contribution can be summarized as follows: 
Firstly, we have created a map matching algorithm that is based on a different principle than the state-of-the-art map matching algorithms.
After that, we performed a theoretical analysis of the algorithm and compared its computational complexity to two previously published map matching algorithms. 
Then, we experimentally compared the accuracy and computational speed of all three algorithms on a standard map matching dataset. 

We found that the proposed algorithm has lower computational complexity than the existing approaches and the experimental evaluation has shown that our approach is able to generate the underlying route estimates with the same accuracy as existing methods, but an order of magnitude faster.

\subsection{Related work}\label{sec:related}
Early map matching algorithms, developed before the advent of GPS-enabled smartphones, did not attempt to recover the complete route of the vehicle, but instead, they focused on finding the most likely road segment for each measurement~\cite{white_map_2000}. An incremental algorithm and a global optimization approach based on FrÃ©chet distance were later proposed by Brakatsoulas et al. in~\cite{brakatsoulas_map-matching_2005}.

Newson et al. ~\cite{newson_hidden_2009} formalized map matching using Hidden Markov Model (HMM). In the proposed model, the authors use the network connectivity and measurement error to characterize the emission and transition probabilities of HMM. The HMM-based map matching proved to be robust and accurate. The matching accuracy was shown to be almost perfect even for location measurements with a period of 30 seconds.
A similar solution using different formalization for the global optimization was presented in~\cite{lou_map-matching_2009}, adding temporal consistency as a transition criteria.
A graph search based approach that search for a path in road network that resembles the GPS records was presented in~\cite{wolfson_weight-based_2004}. 
The idea is similar to the approach we propose, however, the article lacks detailed description and performance evaluation against state-of-the-art methods.
Other algorithms focused on high accuracy matching of sparse GPS data 
was presented in~\cite{rahmani_path_2013}.
In~\cite{kuijpers_uncertainty-based_2016}, authors proposed solution that added the k-shortest path algorithm to compute the globally optimal path.
The work in~\cite{tang_estimating_2016} follows a similar direction presenting a time-expanded road network graph as a formalization of the problem.
In~\cite{yin_general_2016}, the authors propose a novel solution that improves the accuracy by incorporating behavioral model in the map matching process. Moreover, they show that trajectory simplification can be used to speed up the map matching process.

For a more comprehensive overview of existing map matching algorithms, we refer the reader to the existing surveys on the topic~\cite{quddus_current_2007, hashemi_critical_2014, kubicka_comparative_2018}.

\section{Problem Statement}

We start by defining the necessary terms: \emph{Road network} is a directed graph, in which the arcs represent the road segments and the nodes represent either the intersections or simply connect two following road segments.
A \emph{location measurement} contains a latitude, longitude, and the time of the measurement. 
Location measurements are typically obtained using a GPS receiver.
A sequence of location measurements that covers a particular period of time is called a \emph{trace}. If we connect the measurements in the trace with line segments, we get a sequence of lines, which we call a \emph{trace linestring}.
When we talk about \emph{ground truth}, we refer to a sequence of consecutive road segments that was driven by the vehicle, while it was collecting location measurements. The ground truth is used to measure the map matching accuracy. 
The \emph{match} is a sequence of consecutive road segments constructed using a map matching algorithm. 

In this paper, we are interested in the map matching problem, i.e., we desire to determine the ground-truth path traversed by a vehicle in a road network from a given trace.

\section{Towards map matching algorithm for large datasets}
In this section, we briefly describe two existing map matching algorithms and discuss their limitations.
A HMM map matching algorithm (HMM-MM) published in~\cite{newson_hidden_2009} is a well-known example of the global map matching algorithm that is both robust and accurate, but also computational heavy, on the other hand, the incremental algorithm from~\cite{brakatsoulas_map-matching_2005} is an example of a local map matching algorithm that is fast, but not as accurate.

\subsection{HMM-MM algorithm}\label{sec:hmm}
First, we consider HMM-based approach by Newson et al.~\cite{newson_hidden_2009}. HMM-based map matching is a global map matching algorithm that leverages Hidden Markov Models to find the most probable path (sequence of road segments) from all possible paths generated by projecting location measurements to the different edges in the road network. The probability of a path is computed using emission probabilities (probabilities of visiting a node) and transition probabilities (probabilities of traveling between specific nodes). In this approach, a measurement noise model is used to determine the emission probabilities, transition probabilities are inferred from the difference between Euclidean and road network distance.

\subsection{Incremental algorithm}\label{sec:incremental}
Second, we consider the local incremental algorithm described in~\cite{brakatsoulas_map-matching_2005}. 
This algorithm first finds the initial edge by choosing it from a set of candidate edges that lie within given threshold distance from the first location measurement. Then, in each step, it tries to match the next measurement to one of the road segments that are connected to the previously matched edge. 
The algorithm can also utilize a look-ahead strategy. I.e., it tries to match a few location measurements ahead and propagates the score back in an attempt to avoid myopic choices that cannot be reversed later on in the process.

\subsection{Limitations of Current Map matching Algorithms}
The simplest local algorithms for map matching exploit only the geometric relation between location measurements and edges~\cite{white_map_2000}. This approach leads to a very unstable match, the selected edges often jump between different roads and directions and the resulting match is often long and chaotic, especially in dense road networks. 

Contrary to the naive approach described above, state of the art algorithms leverage the road network topology to produce better results. Typically, the algorithm consists of two steps: First, for each location measurement, a set of candidate projection on nearby road segments is computed. Then, the algorithm tries to determine the most likely sequence of the candidate projections. 
This process has proven to be effective in producing accurate matches and is still subject of active research. However, the main drawback of this approach is its high computational complexity.
Although the projection candidates can be found efficiently, the complexity of finding the optimal sequence of candidate projection is significant (see Section~\ref{sec:analysis}). %

The incremental algorithm can partially exploit the topological information through the look-ahead procedure, but in our experiments, we found the algorithm has unreliable performance. For sparser traces, the algorithm was often unable to return any solution.

\section{The Graph Search Based Map Matching}
In this section, we present the Graph Search Based Map Matching (GSMM), a precise and fast map matching algorithm.
In contrast to simple geometric algorithms~\cite{white_map_2000}, the incremental algorithm described in Section~\ref{sec:incremental} or the HMM approach (Section~\ref{sec:hmm}), the proposed algorithm does not iterate over the location measurements in the trace. 
Instead, it searches for a path in the road network in a similar manner as Dijkstra's shortest path algorithm. 
The cost of an arc in the search graph encodes the likelihood of the edge being part of the ground truth path given the location measurements. 
The pseudocode of the proposed algorithm is in Algorithm~\ref{alg:main}. 
We start by initializing the priority queue $Q$ and by adding the start node $n_s$ to it with priority 0. Then, we continue by a standard Dijkstra-style loop, during which we add every neighbor $n_n$ of current node $n_c$ into the queue. The algorithm ends when the destination node $n_d$ is reached, or when the queue is empty (which cannot happen unless the network is not strongly connected). 
The critical part of the algorithm is the initialization discussed in Section~\ref{sec:init}, and the node cost computation explained in Section~\ref{sec:costs}.

\begin{algorithm}[t]
add $n_s$ to $Q$ with priority $0$\;
\While{$Q$ is not empty}{		
	$n_c \leftarrow$ head of $Q$\;
    \If{$n_c$ not closed}{
    	set $n_c$ as closed\;
        \If{$n_c = n_d$}{
        	backtrack \;
		}
		\For{$n_n \in$ neighbors of $n_c$}{
        	\If{$n_n$ not closed}{
            	$c \leftarrow computeCost(n_n)$\;
                add $n_n$ to $Q$ with priority $c$\;
            }
        }
	}
}

\caption{\label{alg:main} GSMM algorithm core}
\end{algorithm}

\subsection{Start and end point determination}\label{sec:init}
\begin{figure}
\centering{}\includegraphics[width=0.8\columnwidth]{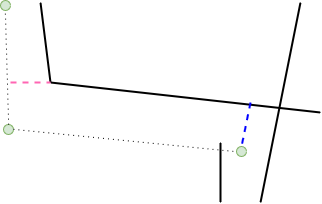}
\caption{\label{fig:init}Initialization example. Green points represent the location measurements, black lines are the roads. The initial edge is chosen such that it minimizes the sum of the distance of the first measurement from the edge (blue dashed line) and the distance between edge endpoint and the trace (pinked dashed line).}
\end{figure}

Determination of start and destination node is the essential part of the algorithm. Although unlike many other algorithms, our solution can recover from a bad choice of the initial node, a mistake in initialization can still affect the final accuracy. 
The initial edge $e_s$ is chosen from the set of edges $E$ as

\begin{equation}
e_s = \argmin_{e \in E}{|r_0, e| + |n_{\mathit{to}}^e, t|}.
\end{equation}
where $r_0$ is the first location measurement, $n_{\mathit{to}}^e$ is the end node of the edge $e$ and $t$ is the trace linestring build by sequentially connecting all location measurements. From the performance reasons, we add to set E only the edges that are closer to $r_0$ than a given threshold. In our experiments, we will use the threshold of 100m. 
When the initial edge is computed, the start node of the edge is chosen as an initial point $n_s$. 
The scenario in Figure~\ref{fig:init} visualizes the costs for the best edge.

The analogical process is performed when determining the destination point. Instead of first measurement $r_0$ we use the last measurement $r_N$, the start node of the edge $n_{\mathit{from}}^e$ and the end node of the best edge is chosen as a destination point.

\subsection{Computing edge costs}\label{sec:costs}
For each neighbor node $n_n$, the cost is computed as a sum of the cost of the current node $n_c$ and the cost of the neighbor $n_n$. The cost of each neighbor $c$ is computed as 
\begin{equation}\label{eq:cost}
c = (c_1 + c_2) l_e / \alpha + c_3,
\end{equation}
where  the cost $c_1$ is the distance of the neighbor node from the trace, $c_1 = |n_n, t|$, and cost $c_2$ is computed as a distance between the middle point of the edge $p_{\mathit{mid}}$ and the trace, $c_2 = |p_{\mathit{mid}}, t|$. These costs have to be multiplied by the length of the edge $l_e$ to normalize for the edge length. The constant $\alpha$ controls the contribution of cost $c_3$ that represents the difference between the length of the edge and distance between the current and  the neighbor node projections on the trace linestring,
\begin{equation}\label{eq:edge_cost}
c_3 = |l_e - l_{\mathit{trace}}|,
\end{equation}where
\begin{equation}\label{eq:l_trace}
l_{\mathit{trace}} = ||p_{{n_c}, t}, p_{{n_n}, t}||_{\mathit{trace}}.
\end{equation}

\begin{figure}
\centering{}\includegraphics[width=0.8\columnwidth]{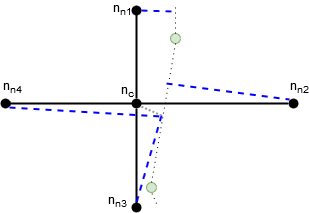}
\caption{\label{fig:cost1}The visualization of the cost $c_1$, depicted as blue dashed line. Note, that the cost for nodes $n_{n_3}$ and $n_{n_4}$ is not the Euclidean distance to the trace, because the neighbors have to be projected after current point. }
\end{figure}

The purpose of these costs can be explained in the series of examples. On Figure~\ref{fig:cost1}, the cost $c_1$ computation is illustrated. An important aspect that can be seen in this figure is that the cost for nodes $n_{n_3}$ and $n_{n_4}$ is not the Euclidean distance to the trace. The reason for this is that the algorithm always project the nodes on the trace chronologically and consequently a neighbor node $n_n$ cannot be projected before the current node $n_c$, that was already projected to the trace in the previous step.

\begin{figure}
\centering{}\includegraphics[width=0.8\columnwidth]{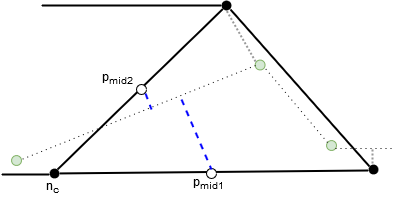}
\caption{\label{fig:cost2}The visualization of cost $c_2$, depicted by the blue dashed line. We can see that the middle point could determine the correct edges even in the case where the incorrect neighbor point has the lowest cost $c_1$ (gray dashed line).}
\end{figure}

The cost $c_2$ is computed analogically to the cost $c_1$, the difference being that we project the middle point of the edge, $p_{\mathit{mid}}$, instead of the neighbor node. The purpose of the cost $c_2$ is illustrated in Figure~\ref{fig:cost2}, where we can see a situation in which relying on the cost $c_1$ would not suffice to determine the correct edge. 

\begin{figure}
\centering{}\includegraphics[width=0.8\columnwidth]{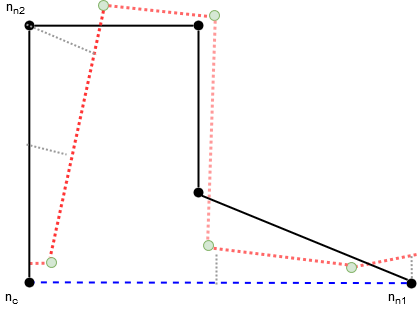}
\caption{\label{fig:cost3}The visualization of the cost $c_3$ for the neighbor node $n_{n_1}$. The final cost is computed as the absolute value of the difference between the edge length (blue dashed line) and the length along the trace (red dotted line). We can see that both cost $c_1$ and cost $c_2$ (gray dashed lines) would mistakenly chose $n_{n_1}$ as the match.}
\end{figure}

We could add more such control points as the costs, but our experiments suggest that this is not effective as there were always cases for which the number of control points was insufficient. Instead, we created another cost $c_3$ that is designed to cover the remaining cases by comparing the length of the edge and the length along the trace between the projection of the current node $n_c$ on the trace (which was fixed in the previous step of the algorithm) and the projection of the node $n_n$. We can see the example where this metric is essential to correctly match the trace in Figure~\ref{fig:cost3}.

\subsection{Candidate projections}
\begin{figure}
\centering{}\includegraphics[width=0.8\columnwidth]{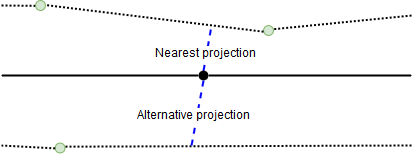}
\caption{\label{fig:projections} The example of ambiguous projection of the node. We can see the nearest projection and one alternative projection on the picture, depicted as the blue dashed line.}
\end{figure}

Sometimes, the vehicle completely changes its direction during a single trip. 
In this case, the projection of a specific edge on the trace linestring can be ambiguous, i.e., multiple points in the trace can have a similar distance to the edge, despite being very far from each other along the trace (see Figure~\ref{fig:projections}). 
We solve such cases by providing alternative projections that are in the same and in the opposite direction of the nearest projection within a distance of 100\,m. Then, we compute the total cost $c$ for each candidate projection and choose the projection with the lowest cost.

\subsection{Forward heuristic cost}
After choosing the best projection for candidate neighbor node we apply another cost $c_4$ to the cost sum, resulting in the equation 
\begin{equation}
c_f = c + c_4,
\end{equation}
where $c$ is the neighbor cost defined above and $c_4$ is computed as
\begin{equation}
c_4 = -\beta ||p_{{n_c}, t}, p_{{n_n}, t}||_{\mathit{trace}}.
\end{equation}
In the above equation, $||p_{{n_c}, t}, p_{{n_n}, t}||_{\mathit{trace}}$ is the distance along the linestring between current and neighbor node projection and $\beta$ is the forward heuristic coefficient. The weight $c_4$ is a heuristic that guides the search and speeds-up the computation, as the edges that lead forward along the trace have a lower cost. By assigning higher values for $\beta$, we can speed up the computation significantly, but higher values can also lead to less accurate results.

\section{Discussion}\label{sec:analysis}
In this section, we discuss the computational complexity, performance, advavantages, and limitations of the three compared algorithms.

\subsection{Complexity}
The worst-case asymptotic complexity of the GSMM algorithm is identical to Dijkstra's algorithm,  $\mathcal{O}(|E| + |N| \cdot \log{|N|})$, where $|E|$ is the number of road segments and $|N|$ is the number of road network nodes. The computational complexity of the HMM-based method is dominated by the Viterbi's algorithm, $\mathcal{O}(|R| \cdot |E|^2)$, where $|R|$ is the number of location measurements in the trace. 
The complexity of the incremental algorithm is $\mathcal{O}(|R|)$. In practice, however, such worst-case complexities are never reached, partly due to the structure of the input and partly due to performance optimization tricks that would be applied in case of practical deployment.  

In HMM map matching, the complexity is reduced by considering only nodes within a fixed radius from the location measurement as candidates -- this drastically reduces the number of edges $|E|$ considered by Viterbi's  algorithm.
In GSMM, the worst case complexity is also virtually never reached, because due to the design of cost function, the search is focused on the vicinity of the trace. 

Finally, real-world traces typically contain only tens to hundreds of location measurements. In result, the time spent on computations that depends on the dataset size $|R|$ can be easily dominated by constant-time computations. This can be easily demonstrated on the incremental algorithm, where the worst-case time complexity is exponential with respect to the lookahead depth $d_l$. For $d_l = 4$, as used in \cite{brakatsoulas_map-matching_2005} and a conservative average branching factor $b = 4$, the complexity of evaluation of each step, which can be expressed as $O(b^{d_l})$, easily dominates the measurement count.

An important difference between the GSMM and other map matching algorithms regarding the complexity is that the complexity of the GSMM is not directly related to the number of location measurements. 
In case of other algorithms, both global and local, the number of algorithm steps is linearly dependent on the number of measurements. 
In result, these algorithms are not suitable for dense traces, sometimes even removing a significant number of measurements from traces is proposed to speed up the algorithm~\cite{yin_general_2016}.

\subsection{Solution quality}
The main shortcoming of the incremental algorithm is that it can choose wrong edges for the match because the lookahead has only constant depth and the edge cannot be changed once chosen for a projection of a location measurement.
In contrast, the HMM-MM and the GSMM search for a match with a globally optimal score.
Yet, the probabilities in HMM-MM and edge costs in GSMM are nothing more than model that uses the sensor noise and network topology.   
In addition, the optimality of the HMM map matching is affected by the fact that it only searches for measurement projection candidates below given distance threshold. 
Without such a threshold, the computation time would be too high for practical application.

\subsection{Advantages of our solution}
\begin{figure}
\centering{}\includegraphics[width=0.8\columnwidth]{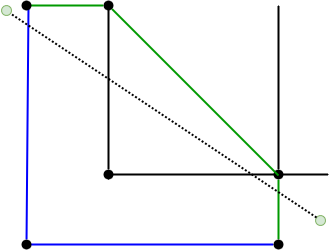}
\caption{\label{fig:beh_example} An example case of a problematic matching related to sparse traces. The green path is closer to the trace, but the ground-truth path is the blue path.}
\end{figure}

As it is clear from the Section~\ref{sec:costs} describing the computation of the weights, GSMM tries to find the path in the road network that is most similar to the vehicle trace. 
This produces correct results most of the time, but in sparse traces, sometimes, the path that is more distant from the trace could be correct as it is demonstrated in Figure~\ref{fig:beh_example}. 
One of the techniques that can be used to match these scenarios correctly is described in~\cite{osogami_map_2013} and~\cite{yin_general_2016}. 
We note that this technique can be incorporated in the GSMM in form of an additional cost. 
Other aspects of human behavior, such as temporal consistency, can be modeled as well, by simply adding the respective costs.

Another advantage of the GSMM is that the edge costs are easy to compute. 
In contrast, in the HMM-MM, the shortest path needs to be computed to evaluate each transitional probability.
The need for such frequent shortest path computation is a significant shortcoming. In fact, we observed that in our implementation, the shortest-path computations account for 98\% of the runtime.

Finally, due to the ability of the GSMM to do fast matching, as it is demonstrated in Section~\ref{sec:results}, we believe that the GSMM can be used for online map matching too, simply by performing the map matching on all location measurements available so far.

\subsection{Limitations of our solution}\label{sec:limitation_our}
Because the GSMM is based on Dijkstra algorithm, it is not able to correctly match traces containing cycles. In practice, we simply break self-intersecting traces in parts and match each part separately.

\section{Benchmark}
In this section, we report on the results of an experimental comparison of the GSMM, the HMM and the incremental approach. 
The algorithms are evaluated using an existing benchmark dataset (see Section~\ref{sec:bench_data}).
To measure solution quality, we use route mismatch fraction (see section~\ref{sec:metric}).  All algorithms and the benchmark were implemented in Python and ran on a desktop PC with Intel Core i5 4950S CPU and 24GB memory.

\subsection{Accuracy Metric}\label{sec:metric}
To measure the map matching accuracy, we used a metric called \emph{Route Mismatch Fraction (RMF)} introduced in \cite{newson_hidden_2009}. The RMF is computed as
\begin{equation}
RMF = \frac{l_{gt}}{l_+ + l_-},
\end{equation}
where $l_{gt}$ is the ground truth length, $l_+$ is the length of the segments in the match that are not in the ground truth (false positive length), and $l_-$ is the length of the segments in ground truth not present in the match (false negative length).

\subsection{Benchmark Dataset}\label{sec:bench_data}
\begin{figure}
\centering{}\includegraphics[width=1\columnwidth]{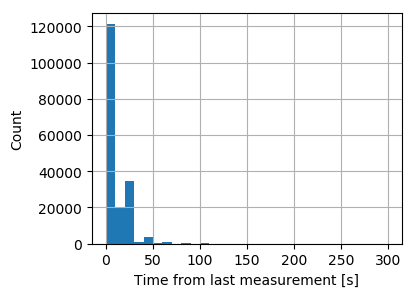}
\caption{\label{fig:delay_hist} Histogram of delay between the traces from Prague taxi trace dataset.}
\end{figure}

As a source of data, we used a previously published dataset\footnote{\url{https://zenodo.org/record/57731}} containing traces together with road networks and ground truths~\cite{kubicka_dataset_2015}. 
The dataset does not contain "clean" traces, but instead, the traces contain artifact like gaps or hives (a large number of measurements in a certain area) that are challenging for map matching algorithms.
The traces in the dataset have the measurement period of one second, longer periods were created by our benchmark by sub-sampling the original trace. Eleven periods between 1 second to 5 minutes length were tested.
The period of 24 seconds represents the average measurement period measured on a real dataset of almost 10 thousand taxi trips in Prague collected by a ride-hailing application Liftago. We measured the average frequency of $0.04$ measurements per second which corresponds to the period of approximately 24 seconds. 
Complete histogram of the delays between points in Prague dataset is in Figure~\ref{fig:delay_hist}.
Because our algorithm cannot handle traces containing cycles, only a subset of traces without cycles was used for evaluation. 

\subsection{Configuration of the evaluated Algorithms}
The HMM-MM algorithm was configured with parameters $\sigma = 50$ and $\beta = 2$, edges in radius of 15 meters was considered as projection candidates. For the incremental algorithm, the parameters was configured as follows: $\mu_d = 10, n_d = 1, a = 0.17, mu_{\alpha} = 10, n_{\alpha} = 7$ The look-ahead depth was set to $4$, start area radius was set to $100$ meters and 
max allowed skipped edges was set to 10.
In the GSMM, the start area radius was set to $100m$ and the forward heuristic cost weight $\beta$ was set to $3$.

\subsection{Results}\label{sec:results}
\begin{figure}
\centering{}\includegraphics[width=1\columnwidth]{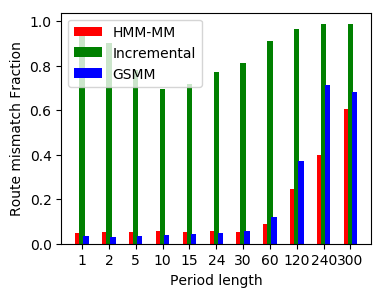}
\caption{\label{fig:accuracy_period} Comparison of the accuracy of the three evaluated algorithms on traces with different sampling frequencies.}
\end{figure}

\begin{figure}
\centering{}\includegraphics[width=1\columnwidth]{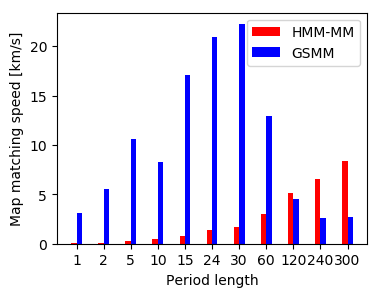}
\caption{\label{fig:time_period} Comparison of speed of the GSMM and the HMM-MM on traces with different sampling frequencies.}
\end{figure}

The comparison of the accuracy of the tested algorithms is in Figure~\ref{fig:accuracy_period}.
As we can see, the accuracy of the incremental algorithm is very low. Thus, we decided to remove incremental algorithm from other comparisons, as it is clearly not able to produce reasonable matches for the benchmark dataset.
The accuracy of the GSMM is comparable to HMM-MM up to the period of 30 seconds, which is more than the average measurement period measured on the real-world dataset (Section~\ref{sec:bench_data}). 
For periods above 30 seconds, the accuracy of both algorithms starts to degrade, with the GSMM declining more rapidly.
Concerning the speed (Figure~\ref{fig:time_period}), the GSMM is faster than the HMM-MM up to the period of two minutes, under one minute period, our algorithm is more than ten times as fast. 

\subsection{Interpretation}\label{sec:discussion}
A first thing to explain is the failure of the incremental algorithm to provide reasonable matches.
We investigated that the algorithm fails on the network areas with very short road segments, where a large number of road segments with similar costs can be assigned to the location measurement. 
We believe that authors of the algorithm measured the accuracy on a simplified road network, where these problems are rare and therefore obtained better results.
The accuracy of GSMM proved to be competitive and in most cases superior to HMM-MM for measurement periods up to 30 seconds. As we show in Figure~\ref{fig:delay_hist}, the available datasets tend to contain most records with a period below 30 seconds. In longer periods, the accuracy of GSMM is declining faster than HMM-MM. 
For traces with a period shorter than one minute, the GSMM is significantly, often an order of magnitude, faster.  
The very important fact is that in global map matching algorithms, the runtime is strongly affected by the number of projection candidates, in the HMM-MM limited by the fixed radius, which was set to $15m$ in the benchmark.
However, in case of a more noisy dataset, this value has to be increased to obtain accurate results.

\section{Conclusion}
The cell phones and connected cars produce large amounts of location data that can be used for various analytical purposes. E.g., this data can be exploited to estimate free-flow speeds or traffic densities in a road network.
A common preprocessing step is map matching, where we desire to infer the path of the vehicle through the road network from sparse and noisy location measurements.

Many map matching algorithms have been proposed that excel in accuracy or robustness with respect to sparse and noisy GPS traces. Yet, with increasing size and availability of location measurement datasets, the processing speed of map matching algorithms becomes important.

We proposed Graph Search based map matching (GSMM) a map matching algorithm based on Dijkstra's algorithm for the shortest path that is focused on reducing the computational complexity. 
We compared the performance of the GSMM together with a state-of-the-art global map matching algorithm (HMM-MM) and one incremental map matching algorithm on a standard, publicly available dataset. 

The results show that for measurements with a period shorter than one minute, our algorithm is up to 10x faster then HMM map matching algorithm and at the same time it achieves higher accuracy than the HMM algorithm.

\printbibliography

\end{document}